\documentclass[showpacs,amsmath,amssymb,aps,prb]{revtex4}
\usepackage{graphicx}

\begin{document}
\def\kag{{\it kagom\'e }}
\title{Numerical Contractor Renormalization applied to strongly correlated systems}

\date{\today}

\author{Sylvain Capponi}
\affiliation{
  Laboratoire de Physique Th\'eorique, CNRS UMR 5152,
  Universit\'e Paul Sabatier, F-31062 Toulouse, France
}

\begin{abstract}
  We demonstrate the utility of effective Hamilonians for studying strongly correlated systems, such as 
  quantum spin systems. 
  After defining local relevant degrees of freedom, 
  the numerical Contractor Renormalization (CORE) 
  method is applied in two steps: 
  (i) building an effective Hamiltonian with longer ranged interactions up to a certain cut-off
  using the CORE algorithm and
  (ii) solving this new model numerically on finite clusters by exact diagonalization and 
  performing finite-size extrapolations to obtain results in the thermodynamic limit. 
  This approach, giving complementary information to analytical treatments of the CORE 
  Hamiltonian, can be used as a semi-quantitative numerical method.
  For ladder type geometries, we explicitely check the accuracy of the effective models by
  increasing the range of the effective interactions until reaching convergence.
  Our results both in the doped and undoped case are in good agreement with 
  previously established results.
  In two dimensions we consider the plaquette lattice and the \kag lattice as non-trivial
  test cases for the numerical CORE method. As it becomes more difficult to extend the range 
  of the effective interactions in two dimensions, we propose diagnostic tools (such
  as the density matrix of the local building block) to ascertain the validity of the basis
  truncation. On the plaquette lattice we have an excellent description 
  of the system in both the disordered and the ordered phases, thereby showing that the CORE
  method is able to resolve quantum phase transitions. On the \kag lattice we find that the
  previously proposed twofold degenerate $S=1/2$ basis can account for a large number of 
  phenomena of the spin $1/2$ \kag system and gives a good starting point to study the doped case.  
\end{abstract}

\maketitle


\section{Introduction}

Low-dimensional quantum magnets are at the heart of current interest in strongly 
correlated electron systems. These systems are driven by strong correlations 
and large quantum fluctuations - especially when frustration comes into play -
and can exhibit various unconventional phases and quantum phase transitions.
Similarly, doping these compounds leads to a rich variety of phases, like superconductivity for instance. 
One of the major difficulties in trying to understand these systems is
that strong correlations often generate highly non trivial low-energy
physics. Not only the groundstate of such models is generally not known
but also the low-energy degrees of freedom can not be identified easily.
Moreover, among the techniques available to investigate these systems,
not many have the required level of generality to provide a systematic
way to derive low-energy effective Hamiltonians.

Recently the Contractor Renormalization (CORE) method has been introduced
by Morningstar and Weinstein~\cite{Morningstar1996}. The key idea of the approach is
to derive an effective Hamiltonian acting on a truncated local basis set, in order 
to exactly reproduce the low energy spectrum. In principle the method is exact 
in the low energy subspace, but only at the expense of having {\it a priori} long-range 
interactions. The method becomes most useful when one can significantly 
truncate a local basis set and still restrict oneself to short-range effective
interactions. This however depends on the system under consideration and has to
be checked systematically.
Since its inception the CORE method has been mostly used as an analytical method 
to study strongly correlated systems \cite{Weinstein2001,Altman2002,Berg2003,Budnik2004}. 
Some first steps in using the CORE approach and related ideas in a numerical framework have
also been undertaken \cite{Piekarewicz1997,Piekarewicz1998b,Capponi2002,Malrieu2001,Al2004b}.

The purpose of the present paper is to 
 explore the numerical CORE method as a 
complementary approach to more analytical CORE procedures (see the contribution by A.~Auerbach in this volume), and to  
discuss its performance in a variety of strongly correlated systems, both 
frustrated and unfrustrated. The approach consists basically of numerical exact
diagonalizations of the effective Hamiltonians. Furthermore we discuss some criteria and tools useful to estimate the
quality of the CORE approach. More technical details can be found in related work done in collaboration 
with D.~Poilblanc~\cite{Capponi2002} and with A. L\"auchli and M. Mambrini~\cite{Capponi2004}. 

After reviewing the CORE algorithm, we will present numerical applications to one-dimensional (1D) systems.
  We will show that the numerical
CORE method is able to get rather accurate estimates of physical properties
 by successively increasing the range of the effective interactions.
Then, we discuss two-dimensional (2D) magnetic  systems. As in 2D a long ranged
cluster expansion of the interactions is difficult to achieve on small clusters, we will 
discuss some techniques to analyze the quality of the basis truncation. We illustrate these issues
on two model systems, the plaquette lattice and the \kag lattice.
The plaquette lattice is of particular interest as it exhibits a quantum phase transition from a 
disordered plaquette state to a long range ordered N\'eel antiferromagnet, which cannot be reached by 
a perturbative approach. We show that a range-two effective model captures many aspects
of the physics over the whole range of parameters.
The \kag lattice on the other hand is a highly frustrated lattice built of corner-sharing triangles and 
it is one best-known candidate
systems for a spin liquid groundstate. A very peculiar property is the exponentially large number of
low-energy singlets in the magnetic gap. We show that already a basic range two CORE approach is able
to devise an effective model which exhibits the same exotic low-energy physics.

\section{Low-energy emerging degrees of freedom}
In various fields,  the high-energy description can be well captured by a well-known model, such as the Hubbard or t-J 
models in the context of high temperature superconductors. However, one is interested in low-energy properties, or similarly 
long-distance behaviour, which are difficult to compute numerically due to system size limitations.

The spirit of Wilson's real-space renormalization group~\cite{Wilson1975} is that one can integrate out local 
degrees of freedom (i.e. high-energy) in order to 
define new emerging degrees of freedom and derive an effective model which will be valid on larger distances. 

The definition of relevant degrees of freedom at a given energy scale is a very deep concept in the sense that one can forget many irrelevant details 
and derive an effective theory. 
For instance, chemists know very well that an atom or a molecule are very powerful concepts, even though they do not exist as fundamental particles.

Now, the question is how do we  identify the relevant ``atoms'' and how do we compute an effective theory ? 
The answer is provided by the CORE algorithm.

\subsection{CORE Algorithm}

The CORE method has been proposed by Morningstar and Weinstein in the
context of general Hamiltonian lattice models~\cite{Morningstar1996}.
Later Weinstein applied this method with success to various spin chain models~\cite{Weinstein2001}.
For a review of the method we refer the reader to these original papers 
and also to a pedagogical article by Altman and Auerbach~\cite{Altman2002} (see also the contribution 
in this volume by A.~Auerbach). 
 Here, we summarize the basic steps before discussing some technical aspects
which are relevant in our numerical approach.

{\it CORE Algorithm~:}
\begin{itemize}
\item 
  Choose a small cluster (e.g. rung, plaquette, triangle, etc) and
  diagonalize it. Keep $M$ suitably chosen low-energy states. 
\item
  Diagonalize the full Hamiltonian $H$ on a connected graph 
  consisting of $N_c$ clusters and obtain its low-energy states $|n\rangle$
  with energies $\varepsilon_n$.
\item
  The eigenstates $|n\rangle$ are projected on the tensor product
  space of the states kept
  and Gram-Schmidt orthonormalized in order to get a basis
  $|\psi_n\rangle$ of dimension $M^{N_c}$.
  As it may happen that some of the eigenstates have zero or
  very small projection, or vanish after the orthogonalization
  it might be necessary to explicitely compute more than just the lowest $M^{N_c}$  
  eigenstates $|n\rangle$.
\item
  Next, the effective Hamiltonian for this graph is built as~: $\displaystyle h_{N_c} = \sum_{n=1}^{M^{N_c}} \varepsilon_n
    |\psi_n\rangle\langle \psi_n|.$
\item 
  The connected range-$N_c$ interactions $h^{\rm conn}_{N_c}$ are determined by
  substracting the contributions of all connected subclusters. 
\item
  Finally, the effective Hamiltonian is given by a cluster expansion
  as
$$
    H^{\mbox{\tiny CORE}}=\sum_i h_i +\sum_{\langle
      ij\rangle}h_{ij}+\sum_{\langle ijk\rangle} h_{ijk} +\cdots
$$
\end{itemize}

This effective Hamiltonian \emph{exactly} reproduces the low-energy 
physics provided the expansion goes to infinity. However, if the interactions 
are short-range in the starting Hamiltonian,
we can expect that these operators will become smaller and smaller, at least
in certain situations.
In the following, we will truncate at range $r$ and verify the convergence in 
several cases. This convergence naturally depends on the number $M$ of low-lying
states that are kept on a basic block. By using the reduced density matrix, we will show a way to determine  how 
``good'' these states are.

Once an effective Hamiltonian has been obtained, it is still
a formidable task to determine its properties. Within the CORE
method different routes have been taken in the past. In their 
pioneering papers, Morningstar and Weinstein have chosen to iteratively 
apply the CORE method 
in order to flow to a fixed point that can be analyzed.
 A different approach has been taken in Refs.~[\onlinecite{Altman2002,Berg2003,Budnik2004}]~: There the 
effective Hamiltonian after one or two iterations has been
analyzed with mean-field like methods and interesting results
have been obtained. Yet another approach - and the one we will
pursue in this paper - consists of a single CORE step to
obtain the effective Hamiltonian, followed by a numerical simulation
thereof. This approach has been explored in a few previous studies
\cite{Piekarewicz1997,Piekarewicz1998b,Capponi2002}. The numerical technique we employ
is the Exact Diagonalization (ED) method based on the Lanczos 
algorithm. This technique has easily access to many observables
and profits from the symmetries and conservation laws in the 
problem, i.e. total momentum and the total $S^z$ component. 

\section{Chain and ladder geometries}\label{sec:ladder}
In this section, we describe results obtained on $S=1/2$ spin chain and ladder systems with 
2 and 4 legs respectively. 

\begin{figure}[!ht]
  \centerline{\includegraphics[width=0.5\linewidth]{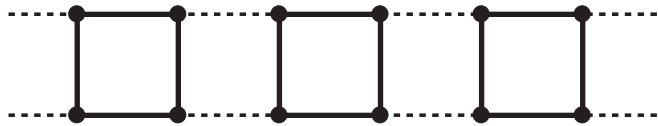}}
  \caption{
     2-leg ladder. The basic block used by CORE is a $2\times 2$ plaquette.
    \label{fig:Lattices1D}
  }
\end{figure}

In the case of doped systems, we use the isotropic t-J model~:
$$ {\cal H}=J\sum_{i,a} \vec{S}_{i,a} \cdot \vec{S}_{i+1,a}
+J \sum_{i,a} \vec{S}_{i,a} \cdot \vec{S}_{i,a+1} - \sum_{i,a} (c_{i,a}^\dagger c_{i+1,a}+h.c.)
- \sum_{i,a} (c_{i,a}^\dagger c_{i,a+1}+h.c.)
$$
that reduces to the usual Heisenberg hamiltonian in the undoped case, 
$\displaystyle {\cal H}=\sum_{\langle ij\rangle} J \vec{S}_i \cdot \vec{S}_j, $
where the exchange constants $J$ will be limited to nearest neighbours.

We have chosen periodic boundary conditions
(PBC) along the chains in order to improve the convergence to the thermodynamic limit. 

\subsection{1D Heisenberg chain}
In this simple example, one is able to iterate the  CORE  procedure in order to obtain the ground-state energy. Let us recall that this model 
has an exact solution for the ground-state energy $e_0=-\ln 2 +1/4$ and has an infinite correlation length so that a numerical approach on a finite system 
is not obvious. Using CORE and solving up to 12 sites, which is very easy even on a small computer, Weinstein has obtained~\cite{Weinstein2001} a ground-state energy 
with a relative accuracy of 
$10^{-5}$. 

A similar idea consists of increasing the size of the initial block, instead of the range of effective interactions, and this has been applied by Malrieu {\it et al.} to the same system~\cite{Malrieu2001}. Solving numerically up to 22 sites, they have a relative error of $10^{-4}$. 

Being able to obtain such an accuracy on a ground-state energy by solving small systems compared to the infinite correlation length is very encouraging. Therefore, 
we have pursued this approach more systematically on other models. 

\subsection{Two-leg Heisenberg ladder}
\label{sec:LadderGeometries}
The 2-leg Heisenberg ladder has been intensively studied and is known 
to exhibit a spin gap for all couplings~\cite{Barnes1993,White1994,Frischmuth1996,Dagotto1996}. 

In order to apply our algorithm, we select a $2\times 2$ plaquette
as the basic unit (see Fig.~\ref{fig:Lattices1D}). The truncated subspace is 
formed by the singlet ground-state (GS) and the lowest triplet state.
Using the same CORE approach, 
Piekarewicz and Shepard have shown that quantitative results can be obtained 
within this restricted subspace~\cite{Piekarewicz1997}. 

\begin{figure}[!ht]
  \includegraphics[width=0.65\linewidth]{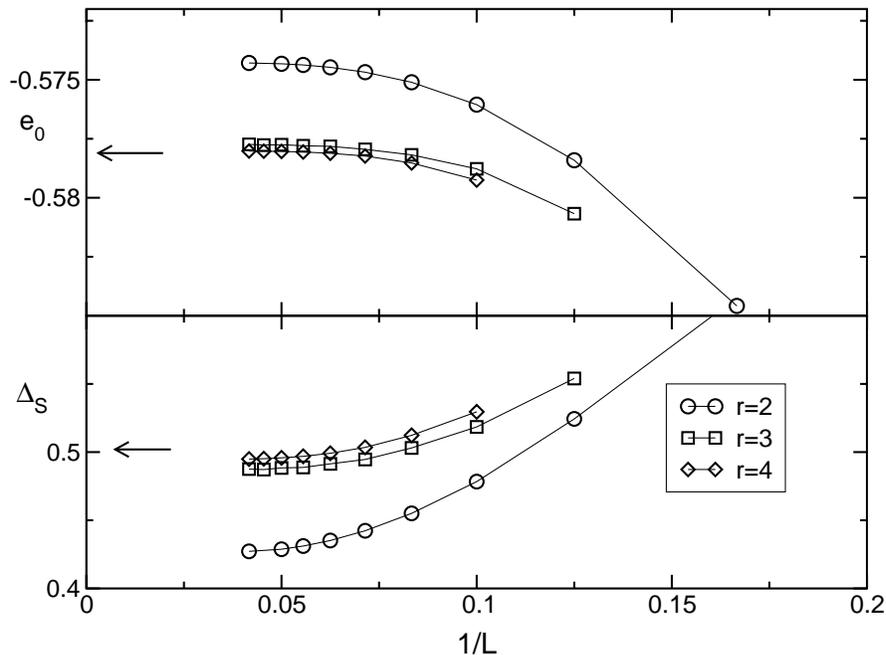}
  \caption{
    Ground-state energy per site ($e_0$) and spin
    gap ($\Delta_S$) of a $2\times L$ Heisenberg ladder using CORE method with
    various range $r$ using PBC. 
    For comparison, we plot the best known extrapolations~\protect{\cite{White1994}} with arrows.
    \label{fig:heisen_2xL}
  }
\end{figure}

Since we are dealing with a simple system, we can compute the effective 
models including rather long-range interactions. It is desirable to compute long-range 
effective interactions since we wish to check how the truncation affect 
the physical results and how the convergence is reached. 

In a second step, for each of these effective models, we perform a 
standard Exact Diagonalization (ED) using the Lanczos algorithm on finite clusters up to $N_c=12$ 
clusters ($N=48$ sites for the original model).
The GS energy and the spin gap are shown in Fig.~\ref{fig:heisen_2xL}. 
The use of PBC allows to considerably reduce 
finite-size effects since we have an exponential convergence as a
function of inverse length.  
CORE results are in perfect agreement with known results and 
the successive approximations converge uniformly to
the exact results. For instance, the relative errors of range-4 results are 
$10^{-4}$ for the GS energy and $10^{-2}$ for the spin gap. This fast convergence is probably 
due to the rather short correlation length in an isotropic ladder (typically 3 to 4 lattice 
spacings~\cite{Greven1996}). 

\subsection{Doped case}
In order to apply CORE, we choose again a $2\times 2$ plaquette as the basic block. 
In addition to the magnetic states, we decide to keep the lowest 2-hole state. 
Therefore, the effective degrees of freedom are hard-core bosons (triplets and hole pairs). A similar 
approach has been used to study the 2-dimensional case~\cite{Altman2002,Chen2004}. 


In Ref.~[\onlinecite{Capponi2002}], we have shown that this effective bosonic model reproduces many features of the doped 2-leg 
ladder such as the persistence of the spin gap, the existence of a triplet-hole pair bound state, as well as the 
characteristic exponent of the superconducting correlations.  
A similar model had been proposed previously~\cite{Siller2001}, but the parameters were obtained from DMRG data 
obtained on large systems. Here, we can deduce the effective parameters by using CORE method, i.e. by solving small clusters. 

\begin{figure}[!ht]
\includegraphics[width=\textwidth]{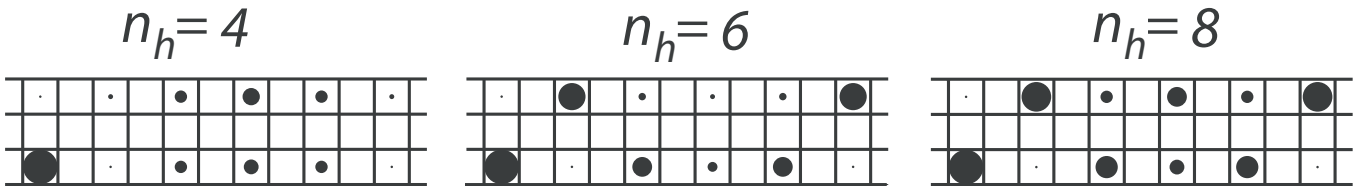}
\caption{Hole-pair density-density correlation on a $4 \times 12$ ladder
at $J/t=0.35$ for different number of holes $n_h$. 
Correlations are measured from the reference plaquette on the
lower left corner. The surfaces of the
dots are proportional to the values of the correlations.\label{hole12}} 
\end{figure}

Following a similar approach, we have also studied the 4-leg t-J ladder~\cite{Capponi2002}. Qualitatively, the physics is very 
similar to the 2-leg case, albeit with smaller energy scales. In Fig.~\ref{hole12}, we draw density-density correlations
obtained with the bosonic effective model for various dopings. 
Upon increasing doping,  we observe a clear
tendency of the hole pairs to align along the diagonal $(1,\pm 1)$
directions (for doping larger than 1/8) with a periodicity corresponding to one pair every two
plaquettes, a behavior also reported in DMRG
calculations\cite{White1997,Siller2002} and reminiscent of the picture
of diagonal stripes. In our case,  PBC were used in the leg direction so that stripes formation is intrinsic and
 not due to any boundary effects. 
 
Therefore, with CORE method, we have both the advantage of working
in a reduced subspace and not being limited to the perturbative
regime. Amazingly, we have observed that for a very small effort (solving a small cluster), the effective 
Hamiltonian gives much better results  than perturbation theory. 
It also gives an easier framework to systematically improve the accuracy by including longer range
interactions. 

For these models, the good convergence 
of CORE results may be due to the fact that the GS in the isotropic limit is 
adiabatically connected to the perturbative one. In the following part we will 
therefore study 2D models where a quantum phase transition occurs as one goes from
the perturbative to the isotropic regime.

\section{Two dimensional spin models}
\label{sec:2D}

In this section we would like to discuss the application of the
numerical CORE method to two dimensional quantum spin systems. We 
will present spectra and observables and also discuss a novel 
diagnostic tool - the density matrix of local objects - in order to
justify the truncation of the local state set.

One major problem in two dimension is the more elaborate cluster
expansion appearing in the CORE procedure.  We therefore try
to keep the range of the interactions minimal, but we still demand a
reasonable description of low energy properties of the system.
We will therefore discuss some ways to detect under what circumstances
the short-range approximations fail and why.

\begin{figure}[!ht]
  \centerline{\includegraphics[width=0.7\linewidth]{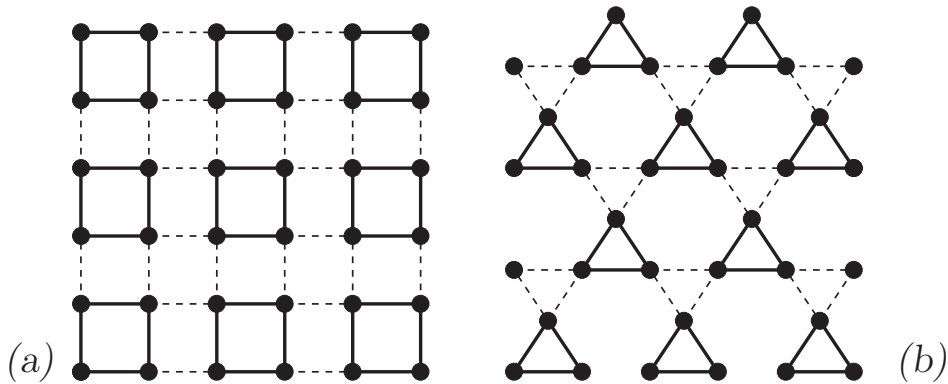}}
  \caption{
    (a) The plaquette lattice. Full lines denote the plaquette bonds $J$, 
    dashed lines denote the inter-plaquette coupling $J'$.
    (b) The trimerized \kag lattice. Full lines denote the up-triangle
    $J$ bonds, dashed lines denote the down-triangle coupling $J'$.
    The standard \kag lattice is recovered for $J'/J=1$.
    \label{fig:Lattices2D}
  }
\end{figure}

As a first example we discuss the plaquette lattice [Fig.~\ref{fig:Lattices2D} (a)],
which exhibits a quantum phase transition from a gapped plaquette-singlet state with
only short ranged order to a  long range ordered antiferromagnetic state as a 
function of the interplaquette coupling
\cite{Koga1999,Koga1999a,Lauchli2002a,Voigt2002}.
We will show that the CORE method works particularly well for this model
by presenting results for the excitation spectra and the order parameter.
It is also a nice example of an application where the CORE method is 
able to correctly describe a quantum phase transition, thus going beyond
a perturbation scheme.

The second test case is the highly frustrated \kag lattice 
[Fig.~\ref{fig:Lattices2D} (b)] which has been 
intensively studied for $S=1/2$ during the last few years 
\cite{Leung1993,Lecheminant1997,Waldtmann1998,Mila1998a,Mambrini2000}.
Its properties are still not entirely understood, but some of the features
are well accepted by now: There is no simple local order parameter detectable, 
neither spin order nor valence bond crystal order. There is probably a small spin
gap present and most strikingly an exponentially growing number of low-energy singlets 
emerges below the spin gap.
We will discuss a convenient CORE basis truncation which has emerged from a 
perturbative point of view~\cite{Subrahmanyam1995,Mila1998a,Raghu2000}.

\subsection{Plaquette lattice}

The CORE approach starts by choosing a suitable decomposition
of the lattice and a subsequent local basis truncation. In the
plaquette lattice the natural decomposition is directly given by 
the uncoupled plaquettes. Among the 16 states of an isolated plaquette
we retain the lowest singlet [$K=(0,0)$] and the lowest triplet
[$K=(\pi,\pi)$]. The standard argument for keeping these states
relies on the fact that they are the lowest energy states in the 
spectrum of an isolated plaquette. 

\begin{figure}[!ht]
\begin{minipage}{0.49\textwidth}
  \centerline{\includegraphics[width=\linewidth]{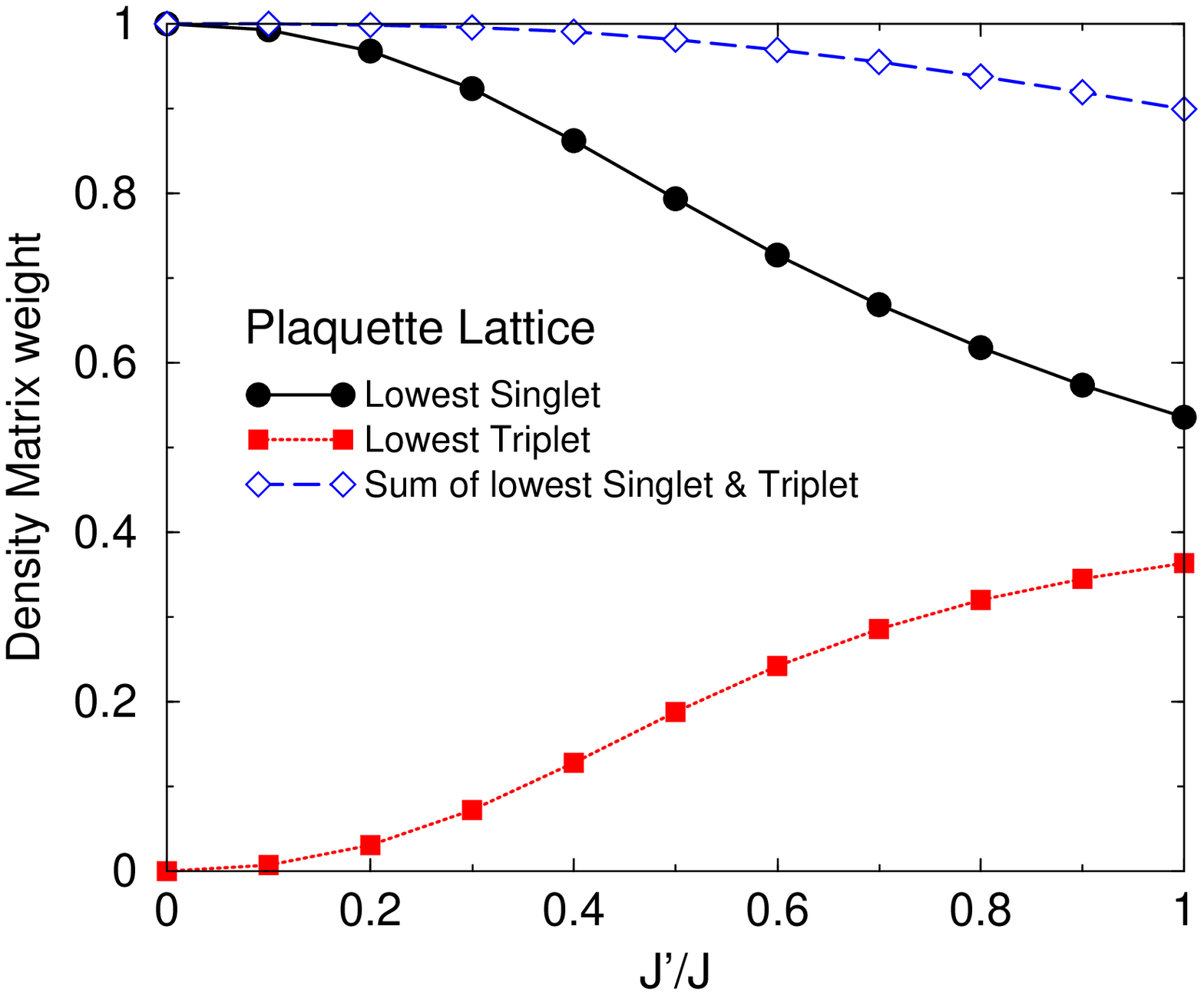}}
\end{minipage}
\begin{minipage}{0.49\textwidth}
  \centerline{\includegraphics[width=\linewidth]{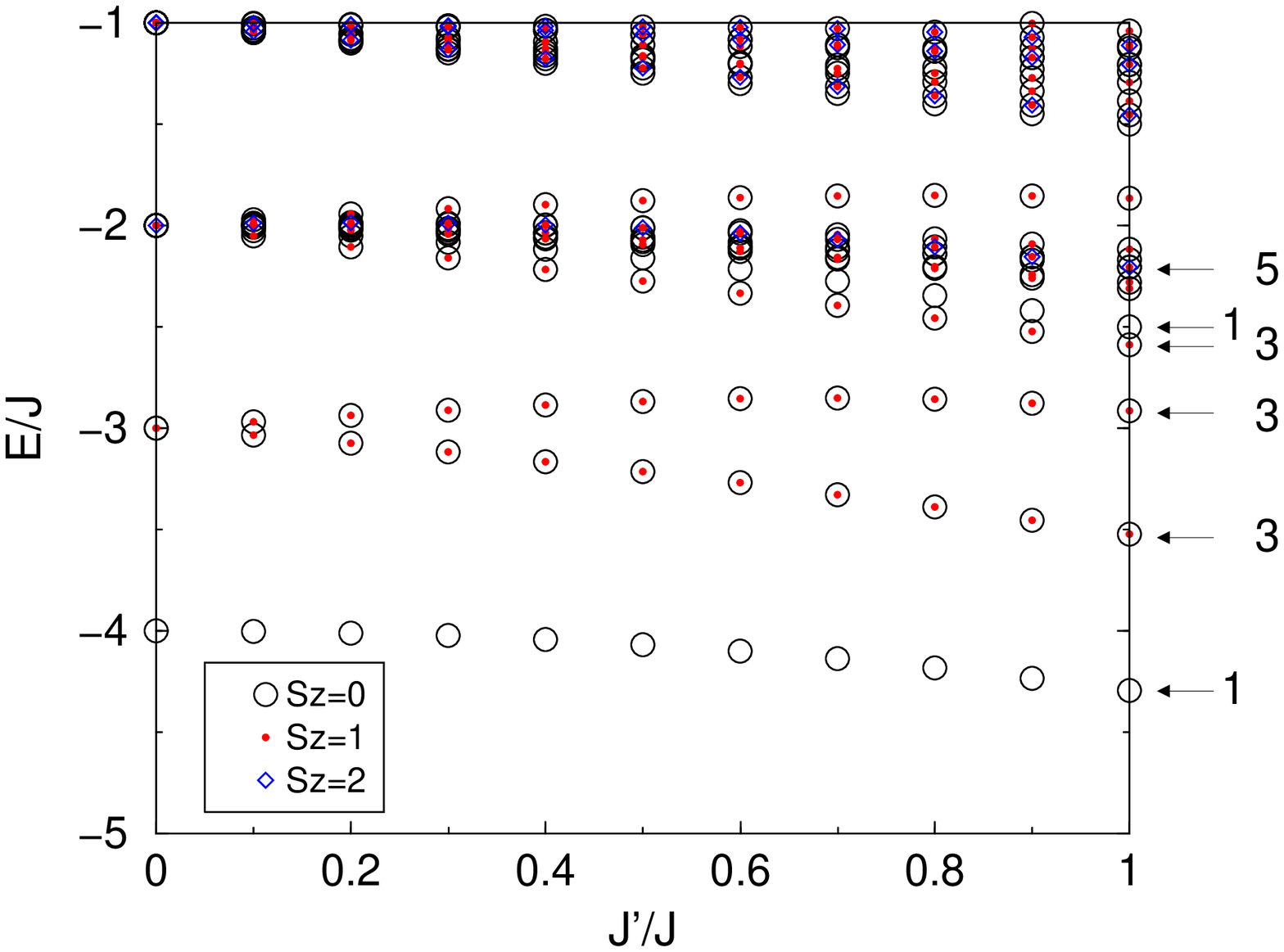}}
\end{minipage}
  \caption{(a) Left panel~: Density matrix weights of the two most important states on
    a strong ($J$-bonds) plaquette as a function of $J'/J$.
    These results were obtained by ED with the original Hamiltonian
    on a $4\times4$ cluster. (b) Right panel~: Low-energy spectrum of two coupled plaquettes. The states 
    targeted by the CORE algorithm are indicated by arrows together
    with their $SU(2)$ degeneracy. 
    \label{fig:Plaq}
  }
\end{figure}

As discussed in Ref.~[\onlinecite{Capponi2004}], the density matrix 
of a plaquette in the fully interacting system gives clear indications
whether the basis is suitably chosen. In Fig.~\ref{fig:Plaq}(a)
we show the evolution of the density matrix weights of the lowest
singlet and triplet as a function of the interplaquette coupling. Even
though the individual weights change significantly, the sum of both
contributions remains above 90\% for all $J'/J\le 1$. We therefore
consider this a suitable choice for a successful CORE application.

A next control step consists in calculating the spectrum of two 
coupled plaquettes, and one monitors which states are targeted by
the CORE algorithm. We show this spectrum in Fig.~\ref{fig:Plaq}(b)
along with the targeted states. We realize that the 16 states of our
tensor product basis cover almost all the low energy levels of the coupled
system. There are only two triplets just below the $S=2$ multiplet which are
missed.


In order to locate the quantum phase transition from the paramagnetic, gapped 
regime to the N\'eel ordered phase, a simple way to determine the onset of long
range order is desireable. We chose to directly couple the order parameter to 
the Hamiltonian and to calculate generalized susceptibilities by deriving the 
energy with respect to the external coupling.  Its simplicity relies on the fact that only
eigenvalues are necessary. Similar approaches have been used so far in ED and QMC 
calculations \cite{Calandra2000,Capriotti2001}.

\begin{figure}[!ht]
  \centerline{\includegraphics[width=0.76\linewidth]{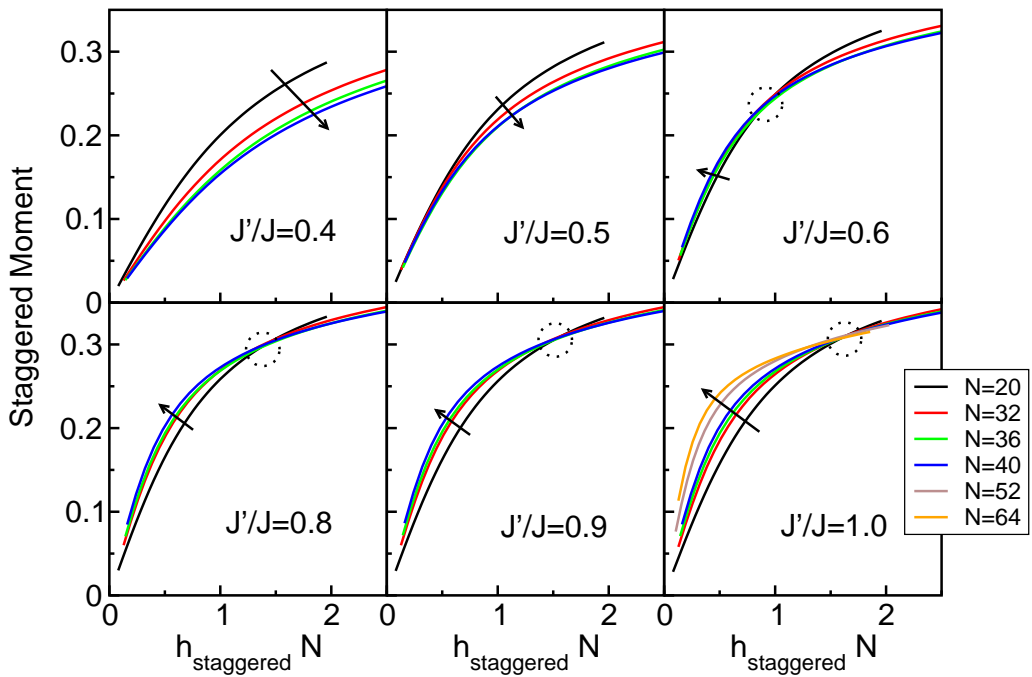}}
  \caption{
    Staggered moment per site as a function of the rescaled applied staggered
    field for the plaquette lattice and different values of $J'/J$.
    Circles denote the approximate crossing point of curves for different system
    sizes. We take the existence of this crossing as a phenomenological 
    indication for the presence of N\'eel LRO. In this way the phase transition
    is detected between $0.5<J'_c/J<0.6$, consistent with previous estimates.
    The arrows indicate curves for increasing system sizes: 20, 32, 36, 40 and also 52, 64
for the isotropic case.
    \label{fig:MstaggeredPlaquette}
  }
\end{figure}

Our results in Fig.~\ref{fig:MstaggeredPlaquette} show the evolution
of the staggered moment per site in a rescaled external staggered field for different
inter-plaquette couplings $J'$ and different system sizes (up to $8\times8$
lattices). We note the appearance of an approximate crossing of the curves for
different system sizes, once N\'eel LRO sets in. This approximate crossing relies
on the fact that the slope of the staggered moment diverges at least linearly in $N$ in the ordered
phase~\cite{Capriotti2001}. We  then consider this crossing feature as an 
indication of the phase transition and obtain a value of the critical point
$J_c/J = 0.55 \pm 0.05$. This estimate is in good agreement with previous studies 
using various methods \cite{Koga1999,Koga1999a,Lauchli2002a,Voigt2002}.

\begin{figure}[!ht]
  \includegraphics[width=0.52\linewidth]{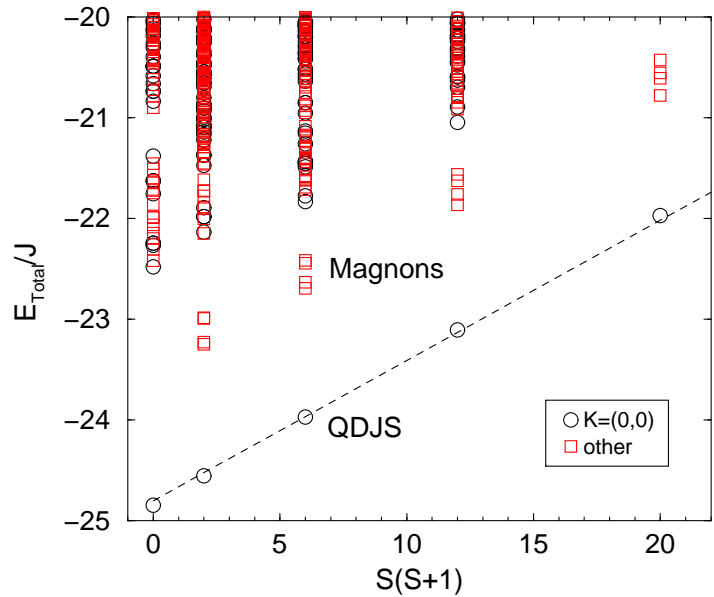}
  \caption{\label{fig:TowerSquare}
    Tower of states obtained with a range-2 CORE Hamiltonian on an
    effective $N=36$ square lattice (9-site CORE cluster)
    in different reduced momentum sectors. The tower of states 
    is clearly separated from the decimated magnons and the rest
    of the spectrum.
  }
\end{figure}

It is well known that the square lattice $(J'/J=1)$ is N\'eel ordered. One 
possibility to detect this order in ED is to calculate the so-called
{\em tower of excitation}, i.e. the complete spectrum as a function
of $S(S+1)$, $S$ being the total spin of an energy level~\cite{Anderson1952}. In the case of
standard collinear N\'eel order a prominent feature is an alignment of the 
lowest level for each $S$ on a straight line, forming a so called
``Quasi-Degenerate Joint States'' (QDJS) ensemble~\cite{Bernu1992},
which is clearly separated from the rest of the spectrum on a finite size 
sample. We have calculated the tower of states within the CORE approach
(Fig.~\ref{fig:TowerSquare}). Due to the truncated Hilbert
space we cannot expect to recover the entire spectrum.
Surprisingly however the CORE tower of states successfully reproduces the general 
features observed in ED calculations of the same model \cite{Sindzingre2003}:
(a) a set of QDJS with the correct degeneracy and quantum numbers 
(in the folded Brillouin zone);
(b) a reduced number of magnon states at intermediate energies, 
both set of states rather well separated from the high energy part of the 
spectrum.

\subsection{\kag systems with half-integer spins}

In the past 10 years many efforts have been devoted to
understand the low energy physics of
the \kag antiferromagnet (KAF) for spins $1/2$ 
\cite{Leung1993,Lecheminant1997,Waldtmann1998,Mila1998a,Mambrini2000}. 
At the theoretical level,
the main motivation comes from the fact that this model is the only known
example of a two-dimensional Heisenberg spin liquid. Even though many questions
remain open, some very exciting low-energy properties of this system have 
emerged. Let us summarize them briefly:
(i) the GS is a singlet ($S=0$) and has no magnetic order. Moreover no
kind of more exotic ordering (dimer-dimer, chiral order, etc.) have been
detected using unbiased methods;
(ii) the first magnetic excitation is a triplet ($S=1$) separated from
the GS by a rather small gap of order $J/20$;
(iii) more surprisingly the spectrum appears as a continuum of states in  all spin sectors.
In particular the spin gap is filled with
an exponential number of singlet excitations: ${\cal N}_{\rm singlets} \sim
1.15^N$;
(iv) the singlet sector of the KAF can be very well reproduced by a
short-range resonating valence bond approach involving only nearest-neighbor dimers.

From this point of view, the spin $1/2$ KAF with its
highly unconventional low-energy physics appears to be a
very sharp test of the CORE method and it was also recently studied in Ref.~[\onlinecite{Budnik2004}].
The case of higher half-integer spins $S=3/2, 5/2 \dots$ KAF is also
of particular interest, since it is covered by approximative experimental
realizations \cite{Limot2002}.

In this section we discuss in detail the range-two CORE Hamiltonians for spin 1/2 
 KAF considered as a set of elementary up-triangles with couplings $J$,
coupled by down-triangles with couplings $J'$ [see Fig.~\ref{fig:Lattices2D}~(b)].

\subsubsection{Choice of the CORE basis}

We decide to keep the two degenerate 
$S=1/2$ doublets on a triangle for the CORE basis. In analogy with the
the plaquette lattice we calculate the density matrix of 
a single triangle embedded in a 12 site \kag lattice, 
in order to get information on the quality of the truncated basis.
The results show that the targeted states exhaust 95\%, which indicates that  the approximation 
seems to work particularly well, thereby
providing independent support for the adequacy of the basis chosen in a related mean-field
study~\cite{Mila1998a}.

\begin{figure}[!ht]
  \centerline{\includegraphics[width=0.6\linewidth]{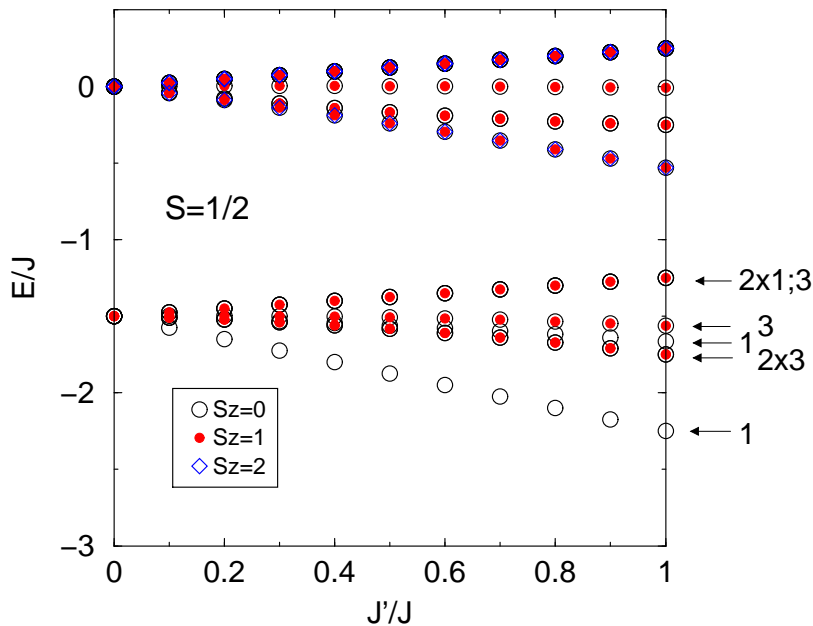}}
  \caption{\label{fig:SpectrumTwoTrianglesOneHalf}
    Spectrum of two coupled triangles in the \kag geometry
    with $S=1/2$ spins. The entire lowest band containing 16 states is
    successfully targeted by the CORE algorithm.
  }
\end{figure}

We continue the analysis of the CORE basis by monitoring the evolution of the 
spectra of two coupled triangles in the \kag geometry 
 as a function of the
inter-triangle coupling $J'$, as well as the states selected by the range-two 
CORE algorithm. The spectrum is shown in 
Fig.~\ref{fig:SpectrumTwoTrianglesOneHalf}. We note the presence of a clear gap 
between the 16 lowest states -- correctly targeted by the CORE algorithm -- 
and the higher lying bands. This can be considered an ideal case for the
CORE method. Based on this and the results of the density matrix we expect the
CORE range-two approximation to work quite well.

\subsubsection{Simulations for $S=1/2$}

We now focus on the effective model describing the standard \kag lattice,
 and we present several distinct physical properties, such as the tower of 
excitations, the evolution of the triplet gap as a function of system size and
the scaling of the number of singlets in the gap. These quantities have been 
discussed in great detail in previous studies of the \kag $S=1/2$ 
antiferromagnet~\cite{Leung1993,Lecheminant1997,Waldtmann1998,Mila1998a,Mambrini2000}. 

\begin{figure}[!ht]
  \centerline{\includegraphics[width=0.62\linewidth]{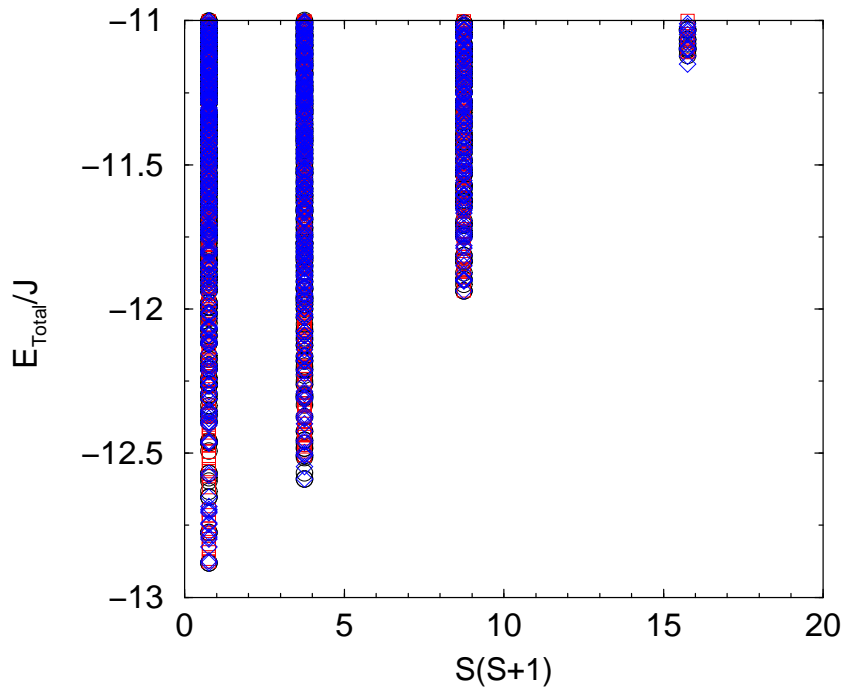}}
  \caption{\label{fig:TowerKagome}
    Tower of states obtained with a range-two CORE Hamiltonian on an
    effective $N=27$ \kag lattice (9-site CORE cluster).
    There is a large number of low-lying states in each $S$ 
    sector. The symbols correspond to different momenta.
  }
\end{figure}

First we calculate the tower of excitations for a \kag $S=1/2$ system on a 27 sites
sample. The data is plotted in Fig.~\ref{fig:TowerKagome}. The structure of
the spectrum follows the exact data of Ref.~[\onlinecite{Lecheminant1997}] rather closely; 
i.e there is no QDJS ensemble visible, a large number of $S=1/2$ states covering
all momenta are found below the first $S=3/2$ excitations and the spectrum is roughly 
bounded from below by a straight line in $S(S+1)$. Note that the tower of states we
obtain here is strikingly different from the one obtained in the N\'eel ordered 
square lattice case, see Fig.~\ref{fig:TowerSquare}.

Next we calculate the spin gap using the range-two and three (containing a closed loop of triangles) CORE Hamiltonians. 
We have a reasonable agreement with ED results when available but there are strong finite size effects.
 The precision of the CORE gap data is
not accurate enough to make a reasonable prediction on the spin gap in the thermodynamic
limit. However we think that the CORE data are compatible with a finite spin gap.

\begin{figure}[!ht]
  \centerline{\includegraphics[width=0.55\linewidth]{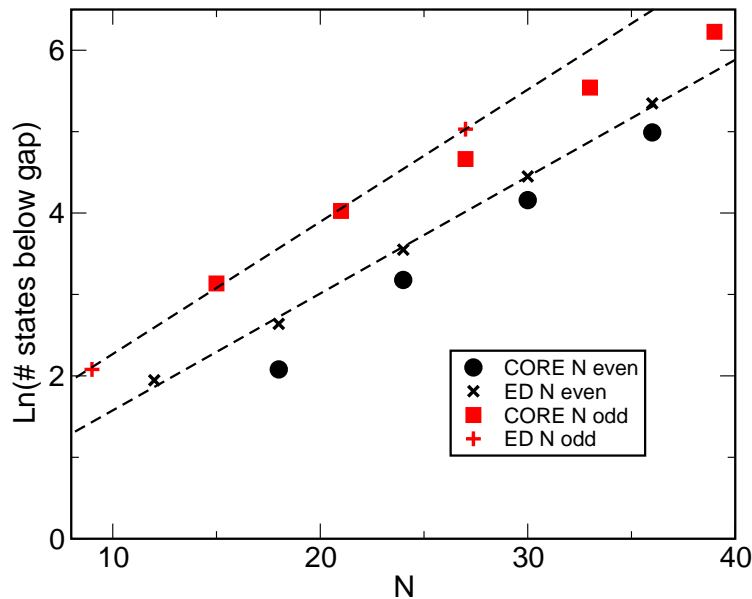}}
  \caption{\label{fig:KagomeOneHalfSinglets}
    Logarithm of the number of states within the magnetic gap. Results obtained 
    with the CORE range-two Hamiltonian. For comparison exact data obtained in Refs.~[\onlinecite{Lecheminant1997,Waldtmann1998}] 
    are shown. The dashed lines are linear fits to the exact diagonalization data.
  }
\end{figure}

Finally we determine the number of nonmagnetic excitations within the magnetic gap
for a variety of system sizes up to 39 sites. Similar studies of this quantity in ED gave 
evidence for an exponentially increasing number of singlets in the gap
\cite{Lecheminant1997,Waldtmann1998}. We display our data in comparison to the exact results
in Fig.~\ref{fig:KagomeOneHalfSinglets}. While the precise numbers are not expected
to be recovered, the general trend is well described with the CORE results. For both even
and odd $N$ samples we see an exponential increase of the number of these nonmagnetic 
states. In the case of $N=39$ for example, we find 506 states below the first magnetic
excitation.

These results emphasize the validity of the two doublet basis for
the CORE approach on the \kag spin 1/2 system. 

It also gives an easier starting point to study the effect of
doping a short-range Resonating Valence Bond state. Using CORE and other techniques, we have shown that the doped
\kag lattice at 1/3-doping undergoes a Peierls transition towards a ``Valence Bond Solid''~\cite{Indergand2005}. This 
instability is due only  to electronic correlations and gives an example of a 2D Bond Order Wave. It  
illustrates how doped antiferromagnets on highly frustrated lattices can partially avoid frustration by lowering
the  lattice symmetry.

\section{Conclusions}

We have discussed the usefulness of real-space renormalization techniques  - the so-called numerical 
Contractor Renormalization (CORE) method - in obtaining local low-energy relevant 
degrees of freedom and an effective theory in the context  of low-dimensional strongly correlated systems. 
This method consists of two steps: (i) building an effective Hamiltonian
acting on the low-energy degrees of freedom of some elementary block; and (ii) studying this new model 
numerically on finite-size clusters, using a standard Exact Diagonalization or similar approach. 

Like in other real-space renormalization techniques, the effective model usually contains longer range
interactions. The numerical CORE procedure will be most efficient provided the effective interactions 
decay sufficiently fast. We discussed the validity of this assumption in several cases. 

For ladder type geometries, we explicitely checked the accuracy of the effective models by
increasing the range of the effective interactions until reaching convergence. Our results on doped and undoped ladders
 are in good agreement 
with previously established results. 

In two dimensions, we have used the density matrix as a tool to check whether the restricted basis gives
a good enough representation of the exact states.  When this is the case, as for the plaquette lattice or
the $S=1/2$ \kag lattice, the lowest order range-two effective Hamiltonian gives semi-quantitative results, 
even away from any perturbative regime. For example we can successfully describe the plaquette lattice,
starting from the decoupled plaquette limit through the quantum phase transition to the N\'eel ordered
state at isotropic coupling. Furthermore we can also reproduce many  aspects of the exotic low-energy
physics of the $S=1/2$ \kag lattice. 

Therefore within the CORE method, we can have both the advantage of working in a strongly reduced subspace 
and not being limited to the perturbative regime.

We thus believe that the numerical CORE method can be used systematically to explore possible ways of
generating low-energy effective Hamiltonians starting from stronlgy correlated models. 


\acknowledgments
I thank 
A. L\"auchli, M. Mambrini and  D.~Poilblanc for their contributions to this work. I also thank A.~Auerbach for
introducing me to the CORE algorithm and for many insightful discussions. Finally, I acknowledge 
 fruitful discussions with F.~Alet, J.-P. Malrieu and F. Mila.

\end{document}